# Hard X-ray Photoelectron Momentum Microscopy and Kikuchi-Diffraction on (In,Ga,Mn)As Thin Films

K. Medjanik[1], O. Fedchenko[1], O. Yastrubchak[2,3] J. Sadowski[4,5], M. Sawicki[4], L. Gluba[6,7],

D. Vasilyev[1], S. Babenkov[1], S. Chernov[1], A. Winkelmann[8], H. J. Elmers[1] and G. Schönhense[1]

1    Johannes Gutenberg Universität, Institut für Physik, D-55128 Mainz, Germany
2    V.E. Lashkaryov Institute of Semiconductor Physics, National Academy of
     Sciences of Ukraine, U-03028 Kyiv, Ukraine
3    Ternopil Ivan Puluj National Technical University Software Engineering Department
     56, Ruska Street, Ternopil, 46001, Ukraine
4    Institute of Physics, Polish Academy of Sciences, Aleja Lotnikow 32/46, PL-02668 Warsaw, Poland
5    Department of Physics and Electrical Engineering, Linnaeus University,
     SE-391 82 Kalmar, Sweden
6    Institute of Agrophysics, Polish Academy of Sciences, Doświadczalna 4,
     PL-20-290 Lublin, Poland
7    Institute of Physics, Maria Curie-Sklodowska University in Lublin, Pl. M.
     Curie-Skłodowskiej 1, PL-20-031 Lublin, Poland
8    Academic Centre for Materials and Nanotechnology, AGH University of Science and Technology,
     Mickiewicza 30, PL-30-059 Krakow, Poland

**Abstract**

Recent advances in the brilliance of hard-X-ray beamlines and photoelectron momentum microscopy facilitate bulk valence-band mapping and core-level-resolved hard-X-ray photoelectron diffraction (hXPD) for structural analysis in the same setup. High-quality MBE-grown (In,Ga,Mn)As films represent an ideal testing ground, because of the non-centrosymmetric GaAs crystal structure itself and In and Mn doping concentrations of few percent. Here we present results of *k*-mapping and hXPD for the title compound with 3% In and 2.5 or 5.6% Mn using hard X-ray photons (3 to 5 keV) at beamline P22 at PETRA III (DESY, Hamburg). Numerical processing (difference or ratio images) emphasizes subtle differences of hXPD patterns like the fingerprint-like hXPD-signatures of As and Ga sites. XPD calculations using the Bloch-wave method show a one-to-one correspondence with the measurements. The hXPD results reveal a predominant Ga substitutional site for Mn. Valence band mapping shows that the Fermi energy lies within the valence band and decreases as the Mn concentration increases. The results support the *p-d* Zener model of ferromagnetism in the title compound. In addition to the shift of the Fermi energy, the band splitting increases with increasing Mn content, which we attribute to an increase of many-body correlations with increasing metallicity of the sample.

# 1. Introduction

Gallium arsenide doped with manganese, (Ga,Mn)As, is a prototypical diluted ferromagnetic semiconductor (DFS) [1] which has raised persistent research interest over the last two decades due to unique premises relying on a combination of ferromagnetism and semiconducting properties in one material [2]. The maximum Curie temperature (close to 200 K) [3,4] is remarkably high compared to other DFS materials, yet not high enough for application-ready devices. Nevertheless, (Ga,Mn)As served as a testing ground for a number of spintronic devices exploiting spin orientation of charge carriers. Due to unique possibilities of controlling magnetic properties in DFS in various ways such as light exposure [5], temperature [6], electrostatic gating [7], pressure [8,9], or doping [10], several types of spintronic devices have been proposed and experimentally realized [11].

In spite of numerous studies, the origin of ferromagnetism in (Ga,Mn)As has been debated [12-15]. Essentially, two models have been proposed to explain the origin of ferromagnetic states: the *Zener model* [16,17] proposes hybridization of Mn states with the GaAs host bands while the *impurity model* considers the appearance of localized Mn impurity states above the GaAs valence band maximum [18]. Up to now, there has been no clear answer, which of the proposed models properly describes the ferromagnetic state. Yet, this is an important question for further development of this material class [19]. So far there is no unified theory, which can explain the origin of the transition from a semiconducting state with localized charge carriers to the metallic state with delocalized carriers in ferromagnetic (Ga,Mn)As with increasing Mn content.

Our recent research has confirmed that homogenous layers with high structural perfection of quaternary (Ga,Mn)(Bi,As), (In,Ga,Mn)As, and pentanary (In,Ga,Mn)(Bi,As) DMS can be grown by low-temperature molecular beam epitaxy (MBE) [20,21]. These multicomponent compounds combine the properties of (Ga,Mn)As (ferromagnetism with the relatively high Curie temperature), Ga(Bi,As) (band structure modification) and (In,Ga)As (local strain creation). The bandgap energy can be changed over a wide range, roughly from 0.4 eV to 1.4 eV in quaternary alloy magnetic semiconductors, which cannot be realized by ternary alloys. For example, the band-gap energy, easy magnetization axis, and band structure can be controlled by changing the In content of (In,Ga,Mn)As [22]. The incorporation of a small amount of Bi into the (Ga,Mn)As layers results in a significant increase in the magnitude of magneto-transport effects probably because of increasing spin-orbit interaction [23]. Similar unusual magnetic effects were observed in (In,Ga,Mn)As material, too [24-27]. The combination of these properties leads a new family of materials that will help to first understand and later tailor their application-relevant properties. In particular, the deeper investigation of quaternary alloy magnetic semiconductors will broaden the possibility of bandgap engineering. Moreover, due to the Mn-induced increase of the lattice constant [2] the (Ga,Mn)As ternary alloy grown on GaAs substrates is always under compressive epitaxial strain, which influences its magnetic properties (magnetic anisotropy) [28]. Quaternary (In,Ga,Mn)As can be obtained as strain-free

III-V DMS, e.g. with use of the lattice matched InP substrates [29]. Additionally, an easy tuning between tensile and compressive epitaxial strain can be achieved by appropriate selection of the In content in (In,Ga,Mn)As grown on InP, which makes this quaternary alloy interesting for prototypical spintronic devices requiring fine tuning of magnetic anisotropy. To our knowledge, so far the results of detailed studies of electronic properties of (In,Ga,Mn)As have not been reported.

Angle-resolved photoelectron spectroscopy (ARPES) is the method of choice to determine the electronic structure in dependence on the Mn concentration. The majority of such studies have been performed in the vacuum ultraviolet (VUV) range [30-32]. The high surface sensitivity in the VUV range emphasizes the band structure in the topmost layers, which can be influenced by band bending or by the surface photovoltage effect. Moreover, physical properties such as carrier-concentration level and strain might differ at surfaces from those of the bulk. In particular, the ubiquitously present donor-like surface traps on the GaAs surface deplete the hole concentration by up to an order of magnitude in the topmost 1-2 nm thick layer [33,34]. In turn, the magnetic constitution of the sub-surface region is altered [35], what is detrimental for applications [36]. Therefore, the complementary measurement of the bulk electronic structure of Mn-doped GaAs is indispensable. Hard X-ray ARPES (HARPES) reveals the true bulk electronic structure [37]. Time-of-flight momentum microscopy provides the push in detection efficiency needed to overcome the obstacle of low photoemission cross section in the hard X-ray regime [38].

In addition to the electronic structure, the position of Mn within the GaAs lattice is a central point in the controversial discussion confronting impurity band and valence band model, because it directly touches the question what mediates ferromagnetism. In the past, the Mn position has been predominantly discussed in theoretical work. For the experimental determination of the Mn position one may employ element-selective diffraction. Here, hard X-ray photoelectron diffraction (hXPD) offers the advantage that the position information and the electronic structure is determined in the identical sample region. The present approach of full-field imaging hXPD bears the advantage that a complete diffractogram at a selected core level can be acquired within typically 20 minutes, because no angular scanning or sample rotation is involved.

In this paper, we present HARPES and hXPD results for (In,Ga,Mn)As thin films with different Mn concentration of 2.5 and 5.6%, recorded using time-of-flight (TOF) momentum microscopy at photon energies up to 5.1 keV. Valence band *k*-patterns resulting from direct transitions into free-electron like final states at such high photon energies allow extracting the band dispersions and their changes with Mn content. We employ the concept of tomographic *k*-space mapping, which works equally well in the soft and hard X-ray range [38,39]. X-ray photoelectron diffraction (XPD) patterns with 0.035° resolution are recorded and reveal significant differences for In, Ga, Mn and As core levels. The Kikuchi-type hXPD patterns show a one-to-one correspondence with theoretical calculations, allowing identifying the dopant sites. We find that the presence of Mn not only causes a shift of the electronic bands relative to the Fermi level but also changes the spin-orbit interaction, visible in the band splitting at the Γ point.

## 2. Techniques

### 2.1 Sample preparation

A series of (In,Ga,Mn)As thin-film samples has been prepared by low-temperature molecular-beam epitaxy (MBE) with an In content of 2.5% and a Mn content *x* ranging from 0 (an (In,Ga)As reference layer) to 5.6%. 30 nm thick (In,Ga,Mn)As epitaxial layers have been grown at ~230 °C on GaAs (001) semi-insulating substrates with pre-deposited fully relaxed 0.8 μm thick $In_{0.075}Ga_{0.925}As$ buffer layer. The buffer layer removes the compressive strain in the top (In,Ga,Mn)As layer, as revealed by photoreflectance studies of the fundamental optical properties of (Ga,Mn)As epitaxial films [40,41]. Tailoring the substrate temperature depending on the intentional Mn content has enabled to maximize the concentration of substitutional Mn (i.e. assuming the Ga lattice position in GaAs), $Mn_{sub}$, incorporated into the InGaAs matrix, and to reduce the $As_{Ga}$ (antisites) and $Mn_I$ defect concentration. The MBE growth has been optimized via the $As_2$ flux, i.e. with $As_2$/(Ga+Mn) flux ratio close to the stoichiometric one, as carefully set during the preceding growth of test/calibration samples. For photoemission studies the (In,Ga,Mn)As layers have been grown on *p*-type conductive substrates to prevent charging effects. The high perfection of a 2-dimensional layer-by-layer growth of (In,Ga,Mn)As has been confirmed by RHEED intensity oscillations, usually observed up to the very end of the growth of 30 nm thick films [42].

### 2.2 Characterization

Micro-Raman spectroscopy was employed to estimate the hole densities in the (In,Ga,Mn)As films. The micro-Raman measurements were performed using an inVia Reflex Raman micro-scope (Renishaw) at room temperature with the 514.5 nm argon ion laser line as an excitation source in backscattering configuration from the (001) surfaces of the (In,Ga)As reference and the two (In,Ga,Mn)As films. The results of the micro-Raman scattering spectroscopy reveals that the longitudinal-optical (LO) phonon mode couples with the hole-gas-related plasmon, forming the so-called coupled plasmon–LO phonon mode (CPPM) for (In,Ga,Mn)As. However, we do not observe this feature in the spectra of the reference (In,Ga)As layers. In this case, only a strong LO phonon line and a very weak (symmetry forbidden) transverse-optical (TO) phonon line located around 290 and 260 $cm^{-1}$, respectively, are observed.

From micro-Raman spectra the hole concentrations have been estimated to be about $10^{19}$ $cm^{-3}$ and $10^{20}$ $cm^{-3}$ in the films with 2.5% and 5.6% Mn content, respectively [32, 43]. The obtained results suggest that, similarly to (Ga,Mn)As [20a] the (In,Ga,Mn)As film with 2.5% Mn shows properties of an insulator-like material whereas the film with 5.6% Mn are similar to a metallic one [44].

It is well known that an As excess - mainly in the form of arsenic antisites, $As_{Ga}$ - occurs at low temperature MBE growth conditions. $As_{Ga}$ antisites act as deep double donors and lead to an *n*-type hopping conductivity in LT-GaAs. Similarly, very diluted (In,Ga,Mn)As is *n*-type, too. The threshold Mn concentration between *n* and *p*-type conduction depends on the actual concentration of $As_{Ga}$ donors and therefore on the actual MBE chamber set up and the growth conditions [45]. With further increase of Mn concentration, the CPPM mode starts to dominate the spectra, simultaneously with its energy shifted towards the TO-phonon-line. This is the direct indication of an increasing hole density with increasing *x*, which can be quantified from the full line-shape fitting [45].

Magnetic properties and the Curie temperature ($T_C$) values for the (In,Ga,Mn)As films were determined using both magnetic-field- and temperature-dependent SQUID magnetometry. Customary cut long Si strips facilitate sample support in the magnetometer chamber [46]. All the data presented here have their relevant diamagnetic contributions of the substrate evaluated at room temperature and subtracted adequately.

Our results are summarized in Fig. 1, indicating ferromagnetic behavior with a non-vanishing remnant magnetization and open hysteresis loops *M*(*H*) of the *x* = 5.6% (In,Ga,Mn)As film at low temperatures. Further measurements (not shown here) reveal an in-plane [-110] easy axis, a characteristic feature ubiquitously met in (Ga,Mn)As films [47]. We determine $T_C$ in this sample from the temperature dependence of the remnant magnetization (TRM) measured along this easy direction. The established magnitude of $T_C$ = 30 K is typical for non-annealed (Ga,Mn)As films [20]. A small inflection point at around $T_S \cong$ 23 K and a rounded shape of both *M*(*H*) loops (both are measured below $T_S$) indicate a spin reorientation transition due to a change from the in-plane uniaxial to an in-plane biaxial magnetic anisotropy, taking place at $T_S$ [48]. SQUID magnetometry has revealed a non-ferromagnetic character of the 2.5% Mn doped (In,Ga,Mn)As layers. It is understood that due to the smaller *x* the free hole concentration is too low to induce a long-range ferromagnetic order above 2 K. This suggestion is supported by the micro-Raman measurements, which reveal that the hole concentration in this layer is below $10^{19}$ cm$^{-3}$.

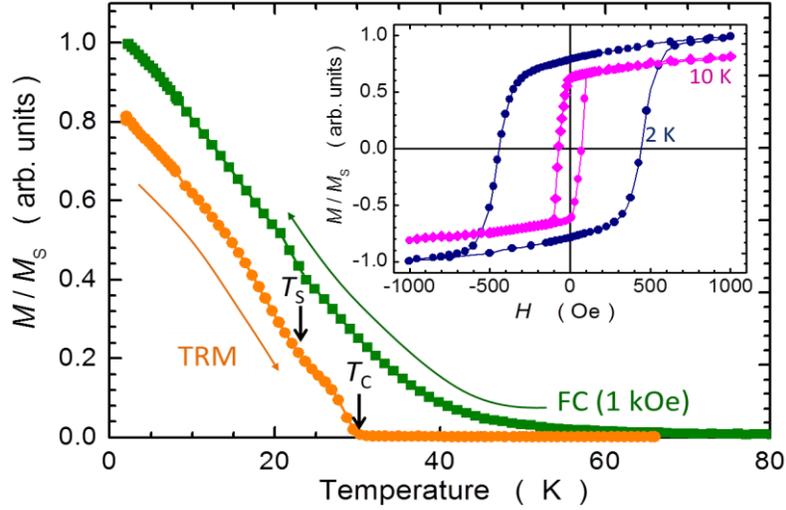

**FIG. 1** (Main panel) Temperature dependence of the magnetization $M$ in an (In,Ga,Mn)As film with 5.6% Mn. The sample is initially field cooled (FC) in a magnetic field of $H$ = 1 kOe (squares) and after quenching the field the thermo-remnant magnetization (TRM) is recorded during warming (circles). The temperature at which TRM vanishes indicates the Curie temperature ($T_C$). $T_S$ indicates the spin reorientation transition between uniaxial (above $T_S$) and biaxial (below $T_S$) in-plane magnetic anisotropies [48]. (Inset) Hysteresis curves at 2 and 10 K. All measurements are performed with $H$ applied in plane along the [-110] uniaxial easy axis. $M$ is expressed with respect to the saturation value $M_S = M$ (2 K, 1 kOe).

## 2.3 Momentum microscopy of bulk valence bands using hard X-rays

The experiment has been performed at beamline P22 [49] of the Synchrotron source Petra III (DESY, Hamburg), belonging to the hard X-ray beamlines with the highest brilliance worldwide. The 3D recording architecture of full-field $k_x$-$k_y$ imaging combined with time-of-flight recording of the kinetic energy $E_{kin}$ overcomes the intensity/resolution problem of hard X-ray angular-resolved photoelectron spectroscopy (HARPES) [50]. At high energies, valence-band momentum distributions are strongly modulated by photoelectron diffraction [51]. However, this modulation can be eliminated by multiplicative correction using core-level diffraction patterns recorded at identical conditions (kinetic energy, $k$-field of view) [38]. The method allows tomographic imaging of the 4D spectral-density function $\rho$ ($E_B$,***k***), tuning the third momentum coordinate $k_z$ by variation of the photon energy [39].

Fig. 2 shows a schematic view of the key elements of the time-of-flight momentum microscope. An objective lens with an electrostatic extractor field between anode and sample yields an achromatic reciprocal image (momentum image) in its backfocal plane (BFP). This momentum image is focused to the image detector by zoom optics via a low-energy drift section, here a delay-line detector (DLD) [52]. This recording mode bears several essential differences in comparison with the conventional ARPES / HARPES approach: The diffractograms are observed in reciprocal space (i.e. on the transversal momentum scale $k_∥$ ) instead of real-space polar coordinates. The *k*-field of view in the present study is up to ~16 Å$^{-1}$ at 7 keV, corresponding to a small polar angular range of 0°–10° recorded with an angular resolution of <0.03°. Larger off-normal observation angles are accessible by polar rotation of the sample using a mode with zero extractor fields [50]. The real-space observation mode (PEEM mode) facilitates inspecting surface quality and easy selection of desired sample areas.

Full-field imaging offers valuable possibilities of image processing, like subtraction or division of different images, a simple way of checking for differences. In order to eliminate effects of electron optics (chromatic aberrations) and to compare identical dynamical diffraction conditions it is essential to adjust the photoelectrons of interest to identical kinetic energies. This is done by tuning the photon energy such that the photoelectrons, whose momentum patterns are to be compared, have the same kinetic energies. In Fig. 2 the momentum pattern of the valence band (VB) and the *3d* core-level diffractograms of As and Ga are shifted to the same energies, controlled by their ToF-spectra (bottom right). This ensures artefact-free ratio images as required for the elimination of the diffraction signature in VB-patterns [38,51] or for a detailed comparison of diffractograms from different atom species in order to determine lattice sites.

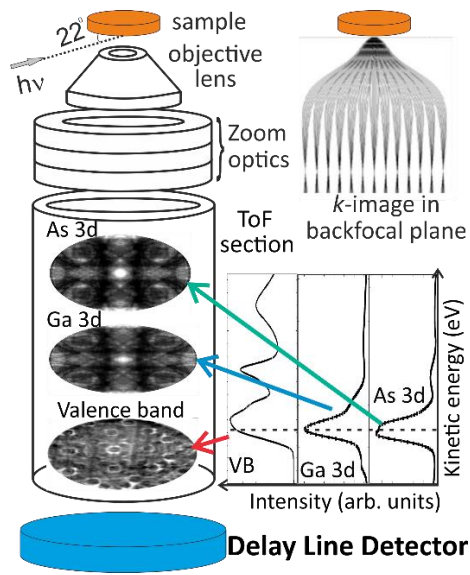

**FIG. 2.** Schematic view of the time-of-flight photoelectron momentum microscope, describing the recording scheme for valence-band studies and core-level hard X-ray photoelectron diffraction (hXPD). The momentum image formed by the objective lens in its backfocal plane BFP is magnified on the ($k_x$,$k_y$,t)-recording delay-line detector (DLD). Energy discrimination is facilitated by a low-energy drift section, which converts the energy distribution to a time spectrum, referenced to the photon-pulse train. The ovals show recorded diffractograms at the As and Ga 3d core-level signals and the momentum pattern of the valence band (after diffraction correction, see below). The corresponding time-of-flight spectra illustrate the principle of adjusting the photoelectrons of interest to identical kinetic energies (dashed line) by proper setting of the photon energy.

## 2.4 Full-field recording of Kikuchi diffraction patterns

Kikuchi diffraction differs from conventional electron diffraction because the source of the diffracted electron wave is located at an emitter atom inside of the material. This process is well known in scanning and transmission electron microscopy (SEM and TEM), where Kikuchi diffraction [53] is observed in terms of electron channeling [54] or electron backscatter diffraction patterns [55]. Quasi-elastic thermal scattering (electron-phonon scattering) of an electron from the primary beam randomizes its *k*-distribution and leads to a new electron wave, emerging from the scattering site. This wave is diffracted at the lattice planes of the crystal before the scattered electrons reach the detector on the vacuum side. The same type of diffraction process happens to a photoelectron from a core level, constituting X-ray photoelectron diffraction (XPD). In both cases the Kikuchi diffraction mechanism is characterized by incoherent, but localized sources of electrons inside a crystal. Because the photoelectrons have element-specific binding energies, XPD allows distinguishing chemically different emitter sites in a crystal. The scattering-induced Kikuchi process in SEM and TEM is not element specific.

In the present work we use full-field photoelectron momentum imaging with hard X-rays as a powerful tool for recording Kikuchi patterns [56]. The special high-energy (> 7 keV) momentum microscope with large *k*-field-of-view (up to 22 Å$^{-1}$ dia.) and high *k*-resolution (0.025 Å$^{-1}$ equivalent to an angular resolution of 0.034° at 7 keV) is ideally suited for fast recording of detailed hXPD patterns with an energy resolution of 0.3-0.5 eV (defined by the photon bandwidth of the monochromator). The key feature of our approach is the possibility to switch between full-field imaging HARPES to full-field high-resolution hXPD, which captures Kikuchi patterns with element selectivity and high site-specificity. Thanks to time-of-flight recording, single counting events are accumulated in I ($E_B$,$k_x$,$k_y$) histogram arrays. Such arrays allow *a-posteriori* analysis of $k_x$-$k_y$ sections at any desired energy or $E_B$-$k_\parallel$ sections in any direction in *k*-space. It is not necessary to know in advance, where in the 4-dimensional ($E_B$,***k***) parameter space the interesting physics is going to happen.

GaAs is a particularly interesting case for photoelectron Kikuchi diffraction because of its non-centrosymmetric structure. We address the question whether this special dissymmetry can be observed by comparing the Kikuchi patterns originating from Ga or As. The site-specific Kikuchi patterns can be used as fingerprints for the location of the dopant atoms. Dopant locations differing from substitutional sites, e.g. tetrahedral and hexagonal interstitial sites, can be identified employing simulated Kikuchi patterns as discussed in the next section.

The measured hXPD patterns are compared with Kikuchi diffraction patterns calculated using the Bloch-wave approach to photoelectron diffraction. The basics of this approach are described in detail in Chapter 2 of [56] for the case of EBSD in the SEM and in [57] for the case of hard X-ray photoelectron diffraction. The possibility to determine point-group resolved orientations of crystals via their Kikuchi patterns has previously been demonstrated using a comparison of simulated and experimental Kikuchi patterns. Examples include the chirality of quartz [58], the mapping of GaP bulk crystals and thin films [59], imaging of inversion twinning in ZnO [60], the determination of the polarity of GaN thin films [60], and the absolute structure mapping of CoSi samples [61].

However, if the electron-scattering factors of the atomic species in the crystal are very similar (which is the case in GaAs), the non-energy-resolved Kikuchi patterns observed in the SEM can become practically centrosymmetric. Winkelmann and Nolze [62] have shown that the limitation of the reduced point-group sensitivity of ECP and EBSD for compounds with atoms of similar electron-scattering strength can be overcome by using energy-dispersive X-ray detection (EDX). EDX provides the element-selective fingerprint required to implement sensitivity on the polarity of an observed crystal surface. This approach works for practically all elements independent on scattering factors. However, it is hampered by the low X-ray fluorescence yield, which is prohibitive for dopant analysis at small concentrations.

In the present study, we employ hXPD of photoelectrons from In, Ga, Mn and As core-levels as a very straightforward way of elemental discrimination. Given the conditions at beamline P22 of PETRA III, high-resolution Kikuchi patterns rich in details are recorded within minutes for intense core-levels and within less than one hour for the weak dopant signals. The one-to-one agreement of experiment and Bloch-wave theory down to the most subtle fine structure near the zone axes [56] makes core-level Kikuchi diffraction a very promising new tool for structural analysis, in particular concerning dopant sites.

The impressive agreement between experiment and Bloch-wave theory for graphite [56] bears an important message: For such materials with low scattering factors [63] (which is also the case for GaAs) and for high energies Bragg reflection at sets of lattice planes is the appropriate model to analyze and interpret hXPD patterns. The computational efficiency of the $k$-space description by the Bloch-wave approach lies in the small number (~100) of Fourier components (corresponding to sets of lattice planes) that need to be considered. It is clear, though, that real-space (i.e. cluster-type) calculations in principle yield identical results if fully converged, as proven by comparative calculations in [57]. Modelling a detail-rich Kikuchi-pattern like those in Figs. 3 and 4 in a cluster-type calculation would demand including ~$10^6$ coherent scatterers, which constitutes a major computational challenge. On the other hand, cluster calculations do not assume long-range periodicity, which is crucial for Kikuchi diffraction. Hence, cluster codes are sensitive to the local arrangement close to the emitter atom and can easily calculate adsorbates on the surface.

Bragg reflection at the system of lattice planes brings out the long-range periodic structure of the material. Nevertheless, core-level Kikuchi diffraction is highly sensitive to the emitter site. We will see below that the calculated patterns for the substitutional and interstitial sites are markedly different, allowing for a clear discrimination. The photoelectron wave emerging from a single dopant atom deep in the bulk feels the periodic lattice of the host material on a length scale of ~20 nm (for the energies we used). The far-field interference pattern sensitively reacts on the location of the emitter atom on a length scale of a fraction of an Å [64].

## 3. Results and discussion

### *3.1 Probing non-centrosymmetry by element-specific core-level Kikuchi diffraction*

Photoelectron diffraction patterns have been recorded at final-state energies of ca. 3.31 and 5.11 keV, the results are shown in Figs. 3 and 4, respectively. The final-state energy is the kinetic energy of the photoelectrons in the crystal (see [39]), being the relevant quantity for the diffraction dynamics. In order to have exactly the same kinetic energies and identical momentum scales for the Ga and As diffractograms, the photon energies are tuned to the values given in the figure captions.

The patterns consist of a pronounced system of Kikuchi bands and lines with a rich fine structure. The experimental and theoretical diffractograms show a one-to-one agreement of the overall structure, the widths of the observed bands and the positions of the lines. Differences occur only in the relative intensities. Eye-catching features are the vertical and horizontal Kikuchi bands of {110} –type, with their bright crossing region in the centre. The width of this band is given by the reciprocal lattice vector $|G_{110}|$ = 3.14 Å$^{-1}$. Another general feature is the system of dark lines throughout the entire pattern. The diameter of the field of view for 5.11 keV (Fig. 4) is 16 Å$^{-1}$, corresponding to a polar angular range of only +/-13°. All patterns exhibit a horizontal and a vertical mirror plane, parallel to the [110] and [1-10] directions.

Figure 3 shows the results for the final-state energy of 3.31 keV. Close inspection of the Ga (a,b) and As (c,d) diffractograms reveals systematic differences, both in the measured and calculated patterns. Arrows with numbers mark characteristic features. As eye-catching signatures, we find two arcs (1) in dark grey, running from left to right for Ga and from top to bottom for As (1'). These arcs (1) and (1') appear in the calculations (b,d) in the same orientations as in the experiment, i.e. rotated by 90° with respect to each other. This is a clear hint on the different orientation of the coordination tetrahedron, see sketch in Fig. 3(g). A second fingerprint are bright horse-shoe-shaped features (2, 2') appearing in the four diagonal directions which touch the arcs close to the rim of the field-of-view. The horse shoes are oriented close to horizontal for Ga (a,b) and close to vertical for As (c,d). Another characteristic are the faint dark crosses (3,3') in the centre of the diffractograms (a-d). The larger angle >90° (marked) is on top / bottom for Ga (a,b) and left / right side for As (c,d). We find an excellent agreement between experiment and theory. The lower resolution in the experimental patterns might indicate a smaller experimental coherence length due to a smaller inelastic mean free path than assumed for the calculation (cf. Fig. 4 in [56]). Further differences show up in relative intensities, as was found in our previous experiment on graphite (cf. Fig. 6 in [56]). These measurements have been performed for the In- and Mn-doped samples.

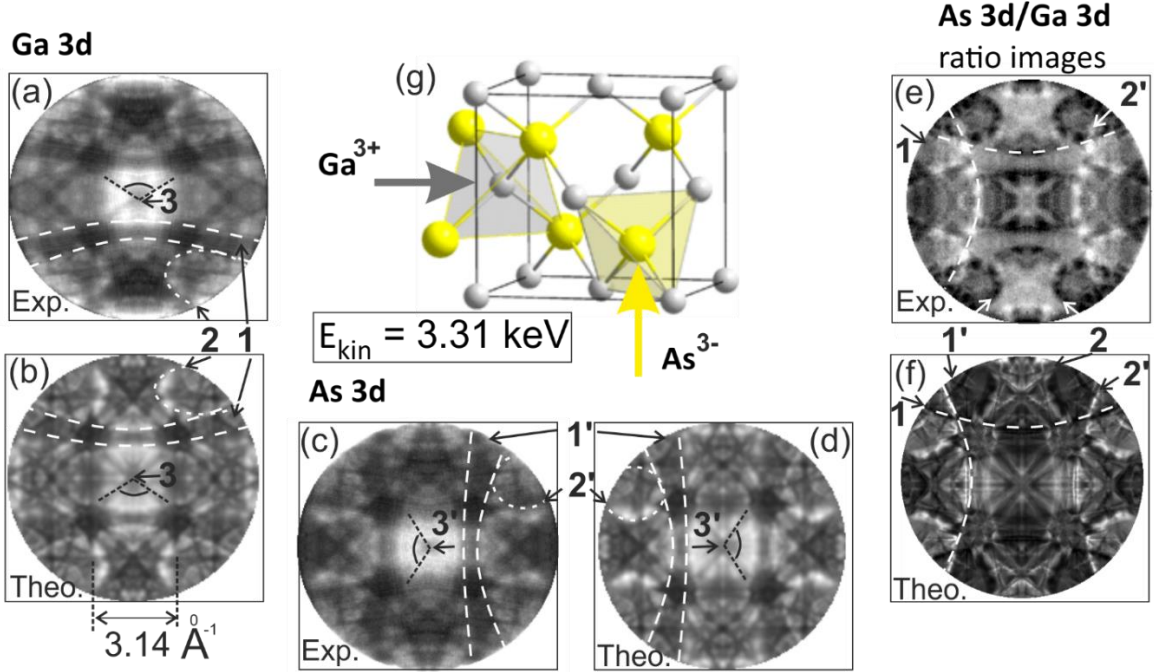

**FIG. 3.** Measured (a,c) and calculated (b,d) diffraction patterns of photoelectrons from the Ga $3d_{5/2}$ and As $3d_{5/2}$ core levels at a kinetic energy of 3.31 keV. Data were recorded for a GaAs (001) thin film doped with 3% In and 2.5% Mn, experimental patterns are two-fold symmetrized. Identical $E_{kin}$ was set by tuning the photon energy to 3317 eV for Ga (a) and to 3340 eV for As (c). (e,f) Ratio images of the As- and Ga-diffractograms, emphasizing the differences originating from the missing inversion center of the zinkblende-type lattice. (g) GaAs unit cell with indicated tetrahedral coordination of the $Ga^{3+}$ and $As^{3-}$ ions. Note that all Ga- and As-tetrahedra are rotated by 90° with respect to each other.

The differences in the diffractograms for As and Ga emitter atoms can be emphasized by means of ratio images, derived via pixel-by-pixel division of the As (c) and Ga patterns (a). Experimental and theoretical ratio images are shown in (e) and (f), respectively. Indeed the characteristic differences are strongly enhanced. In particular the horse-shoe features show up very pronounced in bright (2) and dark dotted lines (2'), dark because the As pattern is in the denominator. Analogously, the arcs appear as rows of bright (1') and dark dots (1) which originate in corresponding rows of dots at the outer side of the arcs in (a-d). Experimental (e) and theoretical ratio images (f) agree very well in all main features. Differences in details are mostly due to the lower resolution of the experiment because of pixel noise and different relative intensities.

Figure 4 shows the analogous results for the final-state energy of 5.11 keV. Due to the different photoelectron wavelength the patterns are markedly different from those in Fig. 3. In the Ga-diffractograms (a,b) we recognize sixfold stars (1) above and below the center, bright spots (2) in four horse-shoe shaped features along the diagonal directions, four bright stripes (3) at the left and right rims, and two V-shaped structures (4) with bright spots at their tips (5) left and right from the center. The As-diffractograms (c,d) show four bright spots (2') in differently-oriented horse shoes, four bright stripes (3') at the upper and lower rim and two V-shaped patterns (4') with bright spot (5') above and below the center. As for 3.31 keV, all As features appear rotated by 90° in comparison with Ga. In Fig. 5(c) the sixfold stars are not present; instead, we observe a diamond-shaped field with inner fine structure (6). This diamond feature appears in the calculation (d) as well, but its inner fine structure looks somewhat different.

Figs. 4(e,f) show the As 3d / Ga 3d ratio images. The experimental ratio image (e) shows a rich fine structure of narrow bright and dark lines, which agree perfectly well with the calculated ratio image (f). Note, e.g., the dark (7) and bright crosses (8), as well as the tilted bright (9) and dark lines (10). Crosses (7,8) and tilted lines (9,10) are again rotated by 90° with respect to each other. The bright spots (2) in (a) and (2') in (c) appear as four characteristic dark – bright features (2,2') in (e) and (f).

Ratio images like those in Figs. 4(e) and 5(e) offer a simple way to eliminate experimental artefacts. Inhomogeneities like uneven illumination of the field of view, effects of distortions due to the electron optics or local sensitivity variations of the detector cancel, because these are present in both images (a, c). Hence, many fine details are emphasized, exploiting the high resolution of the microscope. The comparison of Figs. 5(e) and (f) demonstrates again the one-to-one correspondence of all Kikuchi lines. Actually, this result is not surprising because it is analogous to a Laue diffraction pattern, where the spot positions are solely determined by the reciprocal lattice. Here, each Kikuchi line represents a reciprocal lattice vector (corresponding to a set of lattice planes). Laue spots and Kikuchi lines are both manifestations of the metric of the reciprocal lattice.

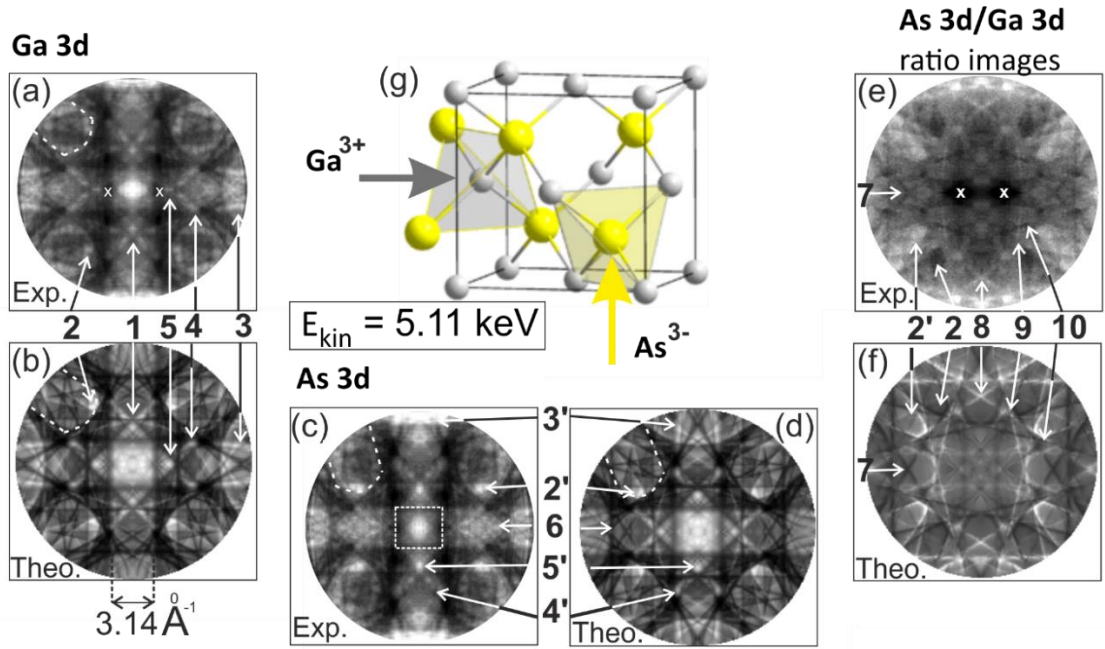

**FIG. 4.** Measured (a,c) and calculated (b,d) Ga $3d_{5/2}$ and As $3d_{5/2}$ core-level photoelectron diffraction patterns at a kinetic energy of 5.11 keV. (e,f) Measured and calculated ratio images of the As- and Ga-diffractograms. Like Fig. 2 for a GaAs (001) thin film doped with 3% In and 2.5% Mn, but for a kinetic energy of 5.11 keV, set by tuning the photon energy to 5117 eV for Ga 3d (a) and to 5140 eV for As (c). The inner part of pattern (c), marked by a square, is shown with different grey level in order to avoid intensity saturation. X marks small dark detector artefacts.

Figure 5 shows a comparison of results taken at identical final-state energies of 3.31 keV and identical settings of the microscope for two samples with different Mn content, 2.5% (top row) and 5.6% (bottom row). Comparison of the diffractograms (a,b,d,e) and ratio images (c,f) shows that all signatures discussed in Fig. 2 are present for both dopant contents. In addition, we have marked the large horse shoes (3, 3') with central cross and dark triangle. The pair of horse shoes is oriented vertically for Ga and horizontally for As.

In the ratio images (c,f) arc 1 from the Ga patterns shows up as chain of bright spots and arc 1' from the As patterns as chain of dark spots. Analogously, the horse shoes 2 and 2' appear bright and dark in (c,f). The elliptical outer shape results from cropping of distorted regions. The patterns in the second row appear slightly more blurred in comparison with the first row. This can be attributed to a higher degree of disorder due to the larger Mn content. Moreover, the central regions in the ratio images (c,f) show significant differences. This may point on increased lattice distortions with increasing Mn content. Distortions on a larger scale in real space show up in the central region in reciprocal space.

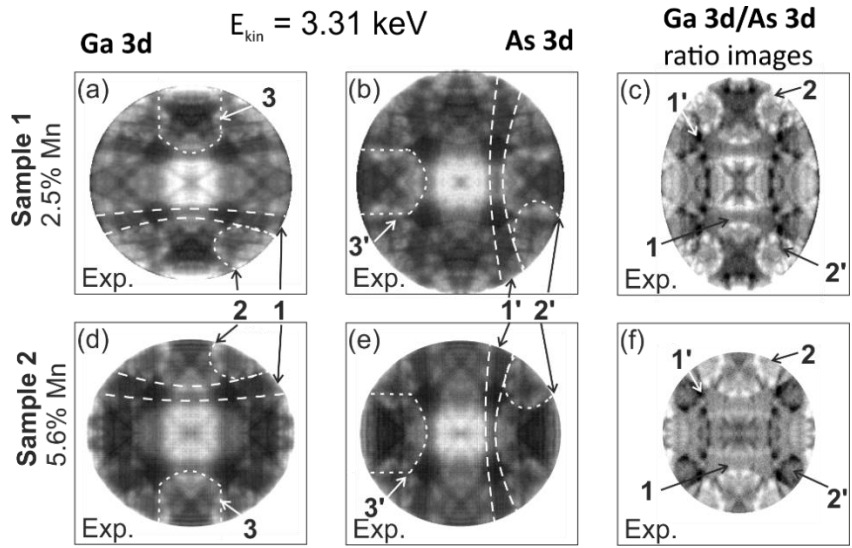

**FIG. 5.** Measured Ga $3d_{5/2}$ and As $3d_{5/2}$ core-level photoelectron diffractograms and corresponding ratio images for sample 1 with 2.5% Mn (a-c) and sample 2 with 5.6% Mn (d-f) for a kinetic energy of 3.31 keV. In the ratio images, outer regions with significant distortions have been cropped, leading to the elliptical shape.

*3.2 Analysis of the lattice sites of the dopants*

The previous section has shown that the Ga and As sites in the GaAs lattice have fingerprint-like signatures of their Kikuchi patterns. The missing inversion symmetry allows an unambiguous assignment of substitutional sites, provided the kinetic energy of the core-level electrons of the dopant is tuned to exactly the same value as used for the reference diffractograms. The question is, however, whether concentrations in the percent range provide sufficient signal for an assignment. The results can be compared with previous work dealing with the location of Mn sites [65,66-68].

Figure 6 shows the diffraction patterns of the four constituents measured at the final-state energy of 3.31 keV. Dotted lines and numbers mark characteristic features. Inspection of the Mn 3d hXPD pattern (c) reveals that the arc (1) runs from left to right and the horse-shoe with center cross is visible on the top and bottom. This is the signature of the Ga substitutional site, in agreement with previous work. Due to the small signal-to-noise ratio of the core-level spectrum, contrast and resolution are reduced in comparison with the main constituents. The pattern had to be smoothed in order to remove statistical noise. The In 3d pattern shows only a very weak contrast, which is not sufficient for an unambiguous assignment of the Ga site.

The second row of Fig. 6 shows calculated Kikuchi diffractograms for Mn atoms on Ga (e) and As sites (f), which are identical to the Ga and As patterns at the same final-state energy. The signatures are in accordance with the experimental patterns (a,b). They show the same orientation of arcs (1) for Ga and (1') for As, of the small horse-shoes along the diagonal (2,2') and the larger horse shoes with cross (3,3'). Incorporation of Mn on the substitutional Ga sites is also evident from the comparison of (c) and (e).

Since they have been discussed as possible Mn sites, we also calculated the Kikuchi-fingerprints of Mn atoms on tetrahedral and hexagonal interstitial sites. Mn on interstitial sites in the GaAs lattice has been analyzed theoretically by Blinowski and Kacman [69] and Mahadevan and Zunger [65]. Indications for such sites have been found by NEXAFS [70], Rutherford backscattering [66], and electron spin resonance (ESR) [71].

Alongside with the two types of interstitials with tetragonal and hexagonal symmetry in the zincblende lattice, there are two non-equivalent tetrahedral interstitial sites with different Kikuchi patterns. The tetrahedral interstitials can be either coordinated to four As or four Ga atoms. If the dopant atoms would occupy energetically-identical sites statistically, the observed Kikuchi patterns would be the average of the two equivalent sites. Calculations revealed, however, that the tetrahedral interstitial site coordinated by As is more stable than that coordinated by Ga. This would induce a dissymmetry in the occupation number. For this reason, we have calculated the Kikuchi fingerprints of both types of interstitial sites separately. A detailed discussion of the site-specificity of Kikuchi patterns is given in [64].

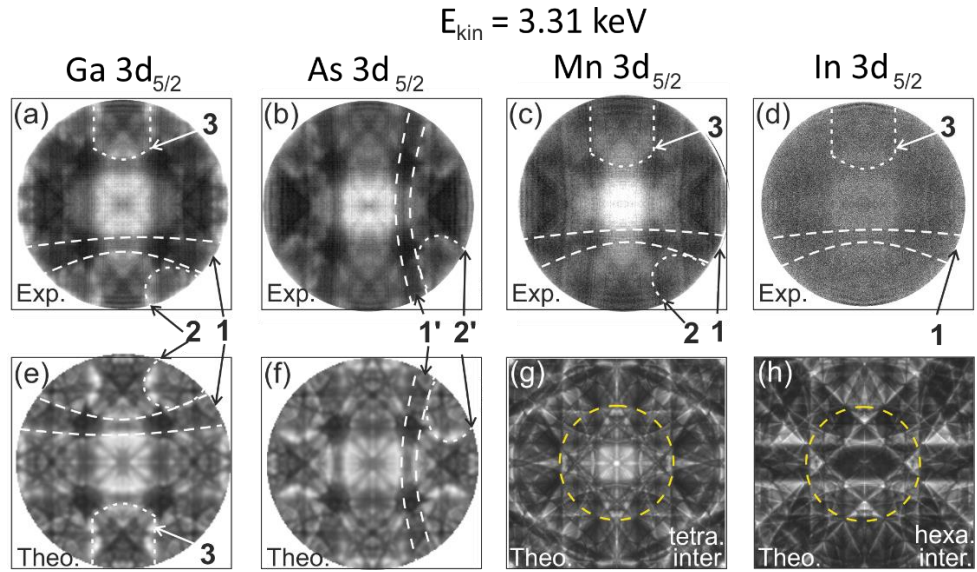

**FIG. 6.** (a-d) Hard X-ray diffraction patterns of photoelectrons from Ga $3d_{5/2}$, As $3d_{5/2}$, Mn $3d_{5/2}$, In $3d_{5/2}$ of the (In,Ga,Mn)As thin film with nominal doping of 3% and 5.6% of In and Mn, respectively. The kinetic energy is set to 3.31 keV by tuning the photon energies to $h\nu$ = 3317 eV, 3340 eV, 3355 eV and 3342 eV for (a, b, c and d). (e-h) Kikuchi diffraction calculation for Mn atoms on Ga (e) and As sites (f) as well as for Mn atoms on tetrahedral (g) and hexagonal interstitial sites (h), all for $E_{kin}$ = 3.31 keV. Dotted lines and numbers mark characteristic features; the dashed circles in (g,h) denote the field-of-view of (a-f).

The calculated hXPD patterns of Mn on tetrahedral and hexagonal interstitial sites are shown in Figs. 6(g) and (h), respectively. The hexagonal site is characterized by dark diagonal Kikuchi bands of {001}-type which cross the horizontal {011}-band. The crossing region appears very dark. This is not compatible with the bright center region of the measured diffractogram (c). The tetrahedral interstitial (g) shows a bright center region but the rest of the pattern does not resemble the experimental pattern (c). For instance, the four large dark spots at the corners of the central bright rectangle in (c) are a prominent feature in the calculation for the Ga site (e) but are very weak in the interstitial calculation (g). From the comparison of experimental and calculated patterns, we conclude that Mn predominantly occupies substitutional Ga sites.

### *3.3 Valence-band mapping by hard X-rays*

Band mapping with hard X-rays is hampered by a strong contribution of valence-band photoelectron diffraction [38,51], which can be largely eliminated by multiplicative correction using a core-level diffraction pattern recorded at the identical kinetic energy. Since the Debye-Waller factor is relevant at such high energies, the patterns have been recorded at a sample temperature of 25 K. Figure 7 shows how this diffraction correction looks like for the title compound. The as-measured valence-band pattern (a) looks very similar to the As 3d core-level diffractogram (b). Note that (a) shows one energy from a 3D data array of several eV width. The energy resolution is defined by the photon bandwidth of 330 meV at this photon energy (3.3 keV).

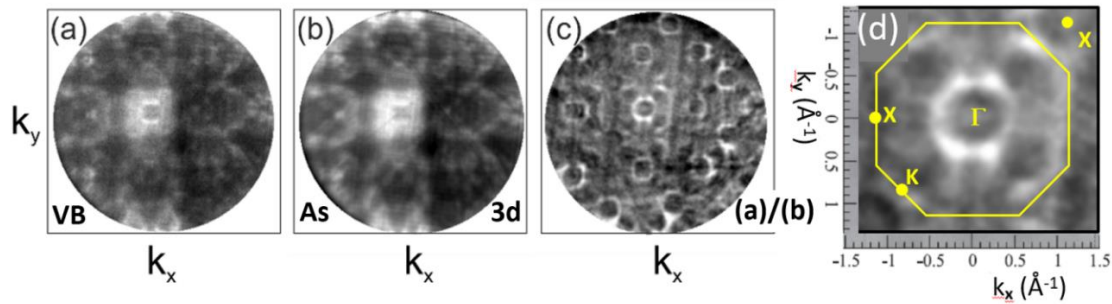

**FIG. 7.** Momentum distributions I ($k_x,k_y$) of photoelectrons from (In,Ga,Mn)As (001) measured at a kinetic energy of 3.3 keV (T = 25K). (a) Constant-energy section of the as-measured valence-band distribution at a binding energy of $E_B$ = 1 eV, dominated by the Kikuchi pattern signature originating from valence-band photoelectron diffraction. (b) Kikuchi pattern of photoelectrons from the As 3d core-level. (c) Valence-band distribution with diffraction eliminated by multiplicative correction, i.e. pixel-by-pixel division of patterns (a) and (b). Now the bulk valence bands in ~12 Brillouin zones (BZ), arranged in a square array are clearly visible. (d) Close-up view of one BZ with marked zone boundary and high-symmetry points Γ, X and K.

In the pixel-by-pixel ratio image (c) the intensity modulation by the Kikuchi pattern is eliminated, uncovering the true valence-band features. The *k*-field-of-view comprises many (~12) Brillouin zones (BZ) in a square array, originating from the section of the final-state energy isosphere with the periodic pattern of [001]-oriented Brillouin zones comprising all band features in 3D *k*-space at the given final-state energy. In valence-band photoelectron diffraction the final-state energy (here 3.3 keV) governs the diffraction dynamics because it defines the photoelectron wavelength. The binding energy (here 1 eV) is not less important, because it defines on which energy iso-surface (one of which is the Fermi surface) the diffraction condition is fulfilled; for details, see [51]. Fig. 7(d) shows a detail of the band features, the central BZ with the contour of the BZ boundary and the high-symmetry points Γ, X and K being marked. The pattern is dominated by an intense band centered around the Γ-point, a weaker band close to Γ and bright features at the periphery of the pattern in the diagonal Γ-K directions.

The *k*-space metric is visible both in the width of the Kikuchi bands in (a,b) and in the distance of equivalent points of the band structure in (c). Given the lattice constant of a = 5.6533 Å, the width of the band is | $G_{110}$ | = 3.14 Å$^{-1}$, and the distance between Γ and X along the image diagonal is |$G_{110}$|/2√2 = 1.11 Å$^{-1}$.

Figure 8 shows the band dispersion as measured for the photon energy of 3300 eV. This energy has been chosen to cut the three-dimensional Brillouin zone in the Γ-X-K plane, assuming a free electron final-state model. Previous work has shown that this approximation is valid at such high energies [39,50]. For both Mn concentrations of 2.5 and 5.6% one observes the valence band dispersions of the heavy hole (HH), light hole (LH), and split-off band (SO) for both high symmetry directions Γ-X and Γ-K. The valence band structure resembles that of undoped GaAs, but there are important differences in the region of the Fermi energy. For comparison, the colored curves in Figs. 8(b,e) show the tight-binding calculation for pristine GaAs from Souma et al. [72], whose energy position has been aligned with the experimental band features.

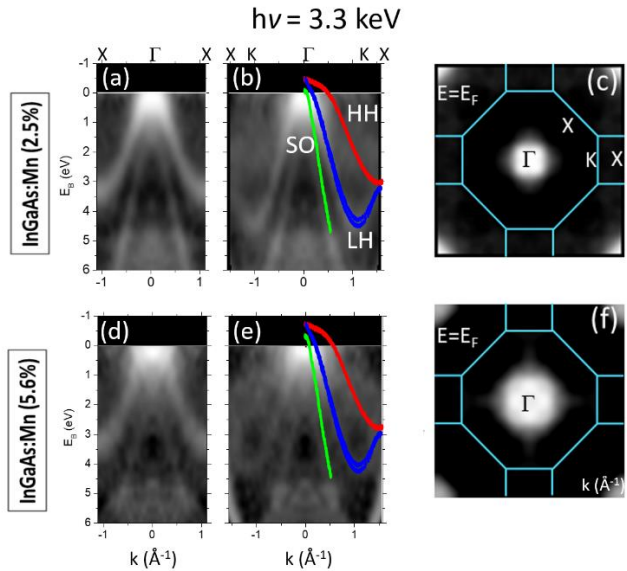

**FIG. 8.** Valence-band dispersion $E_B$-vs-$k_∥$ for (In,Ga,Mn)As thin films with nominal doping of 2.5% Mn in Γ-X (a) and Γ-K direction (b) and for Mn doping of 5.6% along Γ-X (d) and Γ-K (e); photon energy 3.3 keV. (c) and (f) show the corresponding $k_x$-$k_y$ sections at the Fermi energy for 2.5 and 5.6%, respectively. The increased diameter of the bright central area in (f) in comparison with (c) reveals that higher Mn content leads to a decrease of the Fermi energy, giving evidence of acceptor behavior of the Mn atoms. In addition, the splitting between the heavy-hole (HH) and light-hole (LH) bands and the split-off (SO) band increases from 300 meV to 600 meV at the Γ-point. The calculated bands in (b,e) are from [72], their energy position has been adapted to the measured bands.

In contrast to pure GaAs or (In,Ga)As the Fermi energy cuts the heavy and light hole band, indicating metallic behavior (see Fig. 8). The Fermi energy position decreases with increasing Mn concentration. We extrapolate the photoemission data, measured for occupied states, for energies above the Fermi energy using a parabolic approximation. This fitting procedure allows quantifying the shift of the Fermi energy. The maximum of the valence band lies ~100 meV above the Fermi level in the case of 2.5% Mn and ~300 meV for 5.6% Mn. In addition to the shift of $E_F$ with respect to the valence band maximum, we also observe an increase of the binding energy of the split-off band (SO). At the $\Gamma$-point the split-off band lies ~300 meV below the top of the light hole band for 2.5% Mn (a) and ~600 meV for 5.6% Mn content (d).

In classical models, the split-off energy results from a combination of spin-orbit coupling and crystal field. At the $\Gamma$-point the band states are often described in an atomic-like picture, where the HH- and LH-states correspond to hole states with quantum numbers $p_{3/2}|m_j| = 3/2$ and $p_{3/2}|m_j| = 1/2$, respectively. These two states are split by the crystal field. In this picture the SO-state corresponds to a $p_{1/2}$ hole state, which is spherically symmetric. The observed shift with increasing Mn content can thus not be explained in a single electron picture, because the Mn neither increases spin-orbit interaction nor does it substantially change the crystal field. Instead, we attribute the larger splitting to an increase of many-body correlation with increasing metallicity of the sample [73]. The band features in Fig. 8(d,e) appear more noisy than those in Fig. 8(a,b). This is due to increased background intensity with increasing Mn concentration, resulting from scattering of photoelectrons at Mn atoms that act as impurity sites. The same effect was visible in the core-level diffractograms, as discussed in Fig. 5.

Thanks to the high kinetic energy of ~3.3 keV the experiment probes the bulk band structure of the samples with an information depth of about 10 nm. The contribution of the surface region to the total signal is negligible, thus avoiding surface-related effects like the interface band bending caused by surface impurity states. For the high carrier concentrations of the samples investigated here and considering the low temperature of the measurement, the space charge region can extend only 1-2 nm into the bulk of the sample. This information is important for the following discussion of the band positions.

$Mn^{2+}$ ions on Ga sites act as strong acceptors, because their acceptor states are more localized than the hydrogenic-like acceptor states in p-doped GaAs (e.g. Zn-doped). The more localized character results in higher critical carrier density of the metal-insulator transition in Mn doped GaAs (ca. $10^{20}$ cm$^{-3}$). For Mn concentrations beyond the metal-insulator transition (ca. 1.6%) the Mn acceptor levels define the position of the Fermi level, which then shifts below the valence band maximum.

The observed band structure remains largely persistent even for high Mn concentrations, indicating the presence of mobile holes residing in a nearly unperturbed valence band of the GaAs host. These mobile holes are essential for the p-d Zener model of ferromagnetism discussed for diluted magnetic semiconductors. In contrast, the alternatively discussed model for ferromagnetism considers localized Mn impurity states above the GaAs Fermi level, carrying a localized magnetic moment that couples via indirect exchange interaction with the conduction holes. The latter model requires that the Fermi level stays fixed at valence band maximum. This scenario can be excluded by the present measurements, in good agreement with previous work [37,65].

## 4. Summary and conclusions

This paper presents the first combined investigation of the bulk geometric and electronic structure using hard X-rays by the example of the prototypical quaternary diluted magnetic semiconductor (In,Ga,Mn)As. Valence band and core level *k*-distributions are mapped in a full-field imaging time-of-flight momentum microscope optimized for kinetic energies up to 7 keV. The geometric structure is probed by hard X-ray photoelectron diffraction (hXPD), where selected core levels of the constituents are addressed via tuning of the photon energy. hXPD constitutes a new element-specific diffraction method yielding pronounced Kikuchi-type diffractograms with a high site-specificity. Numerical processing (difference or ratio images) emphasizes subtle differences in the Kikuchi patterns like the fingerprint-like signatures of As and Ga sites in the non-centrosymmetric compound. hXPD calculations using the Bloch-wave method show a one-to-one correspondence with the measurements and allow to predict the specific signatures of different dopant sites. For samples with 2.5% and 5.6% Mn content, the Kikuchi patterns confirm that Mn predominantly occupies substitutional Ga sites. The calculated hXPD patterns of Mn on tetragonal and hexagonal interstitial sites in the zincblende-type lattice are not compatible with the measured diffractograms.

The bulk electronic structure of the high-quality MBE-grown (In,Ga,Mn)As films is measured via recording of I ($E_B,k_x,k_y$) 3D data arrays, a highly-efficient approach of hard-X-ray angle-resolved photoelectron spectroscopy (HARPES) [50]. HARPES and hXPD probe identical sample positions and identical depth profiles, thus ensuring that the probed volume is the same for the geometric and electronic information. In previous work, this information required two different experiments, e.g. combining X-ray diffraction or NEXAFS with ARPES at low photon energies. Valence-band and core-level *k*-patterns are recorded at identical settings of the electron optics, allowing for elimination of valence-band XPD. The XPD pattern imprinted on the valence-band signals is essentially identical to the Kikuchi pattern of a core-level signal at the same kinetic energy. Thus, valence-band XPD can be removed by multiplicative correction (pixel-by-pixel division of *k*-images) [38,51].

The paradigmatic high surface sensitivity of photoemission is ideal for experiments, where the surface itself is of interest like in studies of adsorbates or surface chemical reactions. However, when interested in bulk electronic properties the surface sensitivity is a disadvantage, because it strongly emphasizes the topmost few atomic layers. Low-energy ARPES is critical for highly reactive, short-lived surfaces and can even be prohibitive in cases where strong surface states or surface reconstruction are present, or when the band structure is changed in the surface region. Semiconductors constitute a prototypical class of materials with band bending in the surface region. Hence, (In,Ga,Mn)As is an ideal testing ground for the new bulk-sensitive ARPES approach. The non-centrosymmetric compound with zincblende structure and In and Mn doping concentrations of few percent allows to probe the site specificity and sensitivity of photoelectron Kikuchi diffraction.

Our data confirm a predominant Ga substitutional site for Mn up to a concentration of 5.6% without significant contribution of interstitial sites. The HARPES data confirm that in contrast to pure GaAs in this sample the Fermi energy lies within the valence band. This is a consequence of the high concentration of Mn-induced itinerant holes residing in the valence band of the III-V semiconductor host. This result supports the *p-d* Zener model of ferromagnetism in (III,Mn)-V dilute ferromagnetic semiconductors, in accordance with previous results [17,37,72]. We find a decreasing position of the Fermi level with respect to the valence bands as the Mn concentration increases. Moreover, the separation between the split-off band from the heavy- and light-hole bands also increases with increasing Mn content. We attribute this Mn-induced larger band splitting to an increase of many-body correlations with increasing metallicity of the sample.

The observed valence band modifications are fully consistent with the band structure found in previous studies (cf. Fig. 1d in ref. 40 or Fig. 3b in ref. 72), where the Mn-related impurity band is merged with the valence band, forming a metallic-like entity located just above the valence band edge. For weak Mn doping it is known that the band structure remains similar to that of the highly-doped *p*-type (In,Ga)As, where the Moss-Burstein shift of the absorption edge is observed (see Fig. 1c in [40]).

Developing hard X-ray ToF *k*-microscopy further, we are working on refinement towards higher energy resolution and increase of sensitivity by better background suppression. At beamline P22 of PETRA III a resolution of 60 meV FWHM was already demonstrated at a fixed photon energy of 6 keV [50]. A post-monochromator is just being commissioned, which provides this resolution in a wide energy range. Moreover, an integral bandpass filter in the microscope will effectively suppress undesired background electrons. Fast electrons from higher orders of the monochromator/undulator show up in terms of "temporal aliasing" artefacts (for details, see Fig. 5 in [50]). Last not least, lower sample temperatures < 10 K will increase the Debye-Waller factor, thus reducing the background of thermal diffuse scattering in the valence-band momentum patterns [38].

## Acknowledgements

Sincere thanks are due to Christoph Schlueter, Juriy Matveyev and Andrii Gloskovskii (DESY, Hamburg) for excellent support during the beamtimes at beamline P22, PETRA III. We gratefully acknowledge financial support by BMBF (projects 05K19UM1, 05K19UM2) and by Deutsche Forschungsgemeinschaft (Transregio SFB 173 Spin+X 268565370, project A02). Oksana Yastrubchak and Maciej Sawicki are thankful the National Academy of Science of Ukraine and Polish Academy of Science for bilateral cooperation grant.